\newcount\Comments  % 0 suppresses notes to selves in text
\documentclass[%
 twocolumn,
superscriptaddress,
 amsmath,
 amssymb,
 aps, 
 prl,
]{revtex4-2}
\usepackage{chemformula}
\usepackage{float}
\usepackage{graphicx}
\usepackage{dcolumn}% Align table columns on decimal point
\usepackage{bm}% bold math
\usepackage{comment}
\usepackage{mhchem}
\usepackage{todonotes}
\usepackage[normalem]{ulem}
\usepackage{xcolor}
\usepackage{booktabs}
\usepackage{soul}

\newcommand{\kibitz}[2]{\ifnum\Comments=1\textcolor{#1}{#2}\fi}

\begin{document}
\preprint{APS/123-QED}

%\title{\textbf{Dual Polaronic Behavior in Photoexcited \ch{CeO_2}} }% 
\title{Coexisting Large and Small Polarons in Photoexcited \ch{CeO_2}}

\author{V. Mazzotti}
\affiliation{Department of Physics, McGill University, Montréal, Québec H3A 2T8, Canada}
 \author{E. Spurio}%
\affiliation{Istituto Nanoscienze – CNR, S3, Via G. Campi 213/A, 41125 Modena, Italy}
\author{N. Delnour}
\affiliation{Department of Physics, McGill University, Montréal, Québec H3A 2T8, Canada}
\author{M. Hilke}
\affiliation{Department of Physics, McGill University, Montréal, Québec H3A 2T8, Canada}
\author{K. H. Bevan}
\affiliation{Division of Materials Engineering, Faculty of Engineering, McGill University, Montréal, Québec H3A 0C5, Canada}
\author{A. Shluger}
\affiliation{Department of Physics and Astronomy, University College London, Gower Street, London WC1E 6BT, United Kingdom}
\author{P. Luches}
\affiliation{Istituto Nanoscienze – CNR, S3, Via G. Campi 213/A, 41125 Modena, Italy}
\author{D. G. Cooke}
\affiliation{Department of Physics, McGill University, Montréal, Québec H3A 2T8, Canada}
 \email{Contact author: cooke@physics.mcgill.ca}

\date{\today}

\begin{abstract}
Time-resolved terahertz (THz) spectroscopy is used to probe polaron formation in \ch{CeO2}. The ultrafast THz photoconductivity is dominated by contributions from large hole and small electron polarons. Delocalized large hole polarons are formed on a $820\pm50$~fs time scale and are of Fr\"ohlich type, with optical conductivity well described by a simple Drude model and a transient hole mobility of $100~\text{cm}^2/\text{Vs}$. Small electron polarons form via localized coupling to Ce lattice sites, revealed by reduction in the lattice Born effective charge causing a transient phonon softening. These results establish the dual polaron nature of \ce{CeO2} and provide a basis for understanding charge transfer dynamics in metal oxide photocatalysts.
\end{abstract}

\maketitle
\begin{figure}[t!]
    \centering
    \includegraphics[width=\columnwidth]{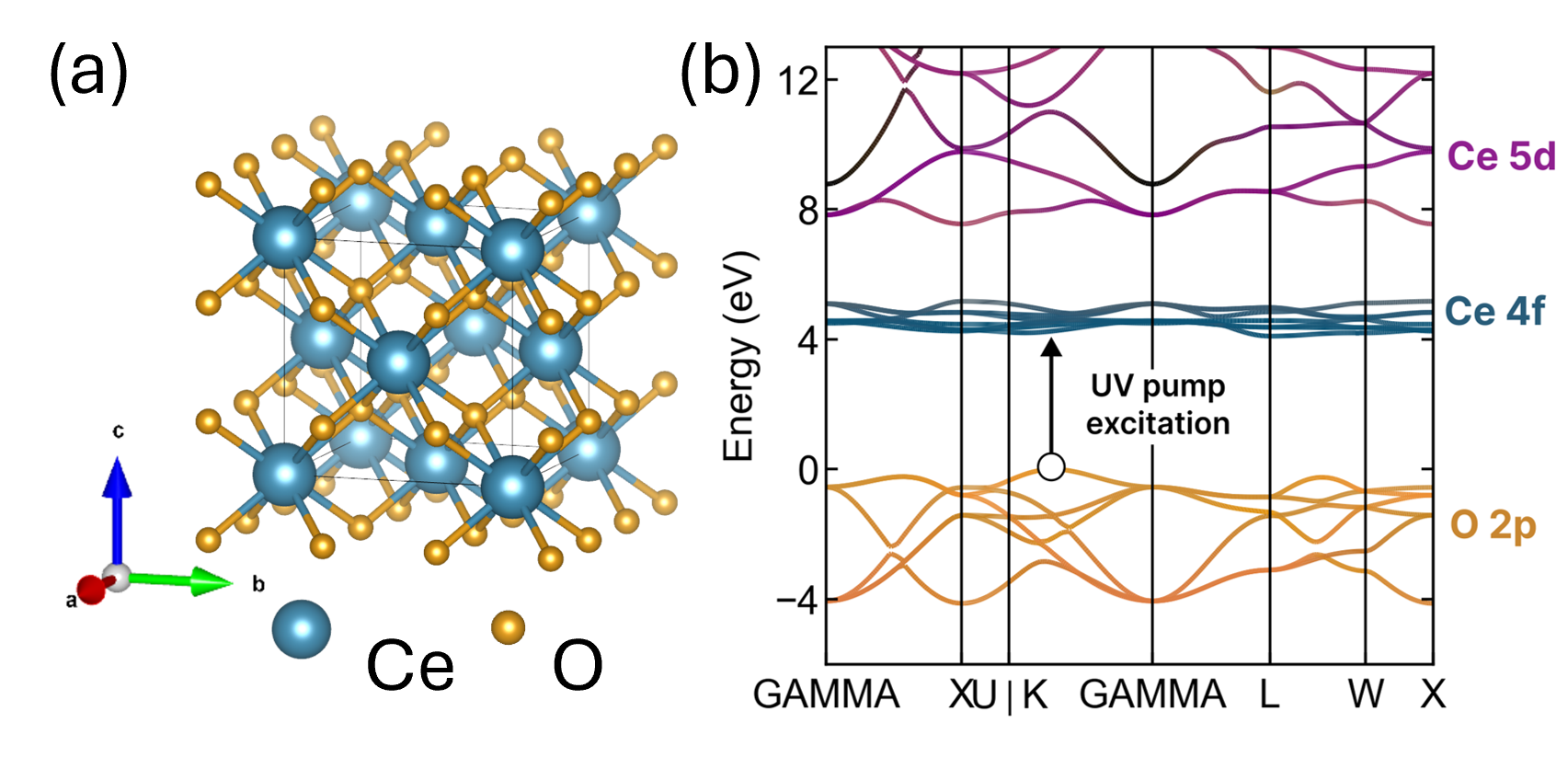}
    \caption{(a)~Fluorite $Fm\bar{3}m$ structure of 
    \ce{CeO2}~\cite{Momma2008}. 
    (b)~Calculated electronic band structure of 
    \ch{CeO2} using the PBE0 hybrid functional, displaying the above-bandgap 
    excitation mechanism via UV (4.64~eV) femtosecond 
    pulses. The excitation promotes electrons from 
    the dispersive O~$2p$ valence band to the 
    localized Ce~$4f$ conduction band, creating 
    mobile holes in the valence band. 
    }
    \label{fig:fig1}
\end{figure}

%\section{\label{sec:intro}Introduction}
Wide-bandgap cerium oxide \ce{CeO2} plays a central 
role in heterogeneous catalysis, energy conversion, and 
oxide electronics~\cite{MontiniChemRev2016}, enabled by 
its remarkable redox flexibility and mixed 
ionic--electronic 
conductivity~\cite{Trovarelli1999,Sun2012}.
Most prominently, \ce{CeO2}  is widely employed as an 
active component in automotive exhaust converters and 
the water--gas shift 
reaction~\cite{Trovarelli1999,MontiniChemRev2016}, 
owing to its exceptional ability to cycle reversibly 
between the \ce{Ce^{4+}} and \ce{Ce^{3+}} oxidation 
states.
This redox flexibility is intimately connected to the 
tendency of electrons to localize on cerium sites as 
small polarons~\cite{FranchiniNatRevMat2021} -- its structure can be seen in Fig.~\ref{fig:fig1}(a). This charge-transfer physics is evident in its electronic structure \cite{ZaanenPRL1985}, with the conduction band heavily dominated by localized \ce{Ce} 4$f$-orbitals and the dispersive valence band by \ce{O} 2$p$-orbitals (see Fig.~\ref{fig:fig1}(b)). 
%The formation of \ce{Ce^{3+}} sites and oxygen defects has been shown to directly affect the photocatalytic performance of \ce{CeO2}~\cite{Choudhury2014}, with the dynamics of charge localization and transport directly relevant to its catalytic function. In this respect, the mobility of electrons and holes in photocatalytic oxides both play a crucial role in ensuring the efficient separation of charge carriers during redox activity \cite{Iqbal2018,Peter2013}.

\begin{figure}[t!]
    \includegraphics[width=\columnwidth]{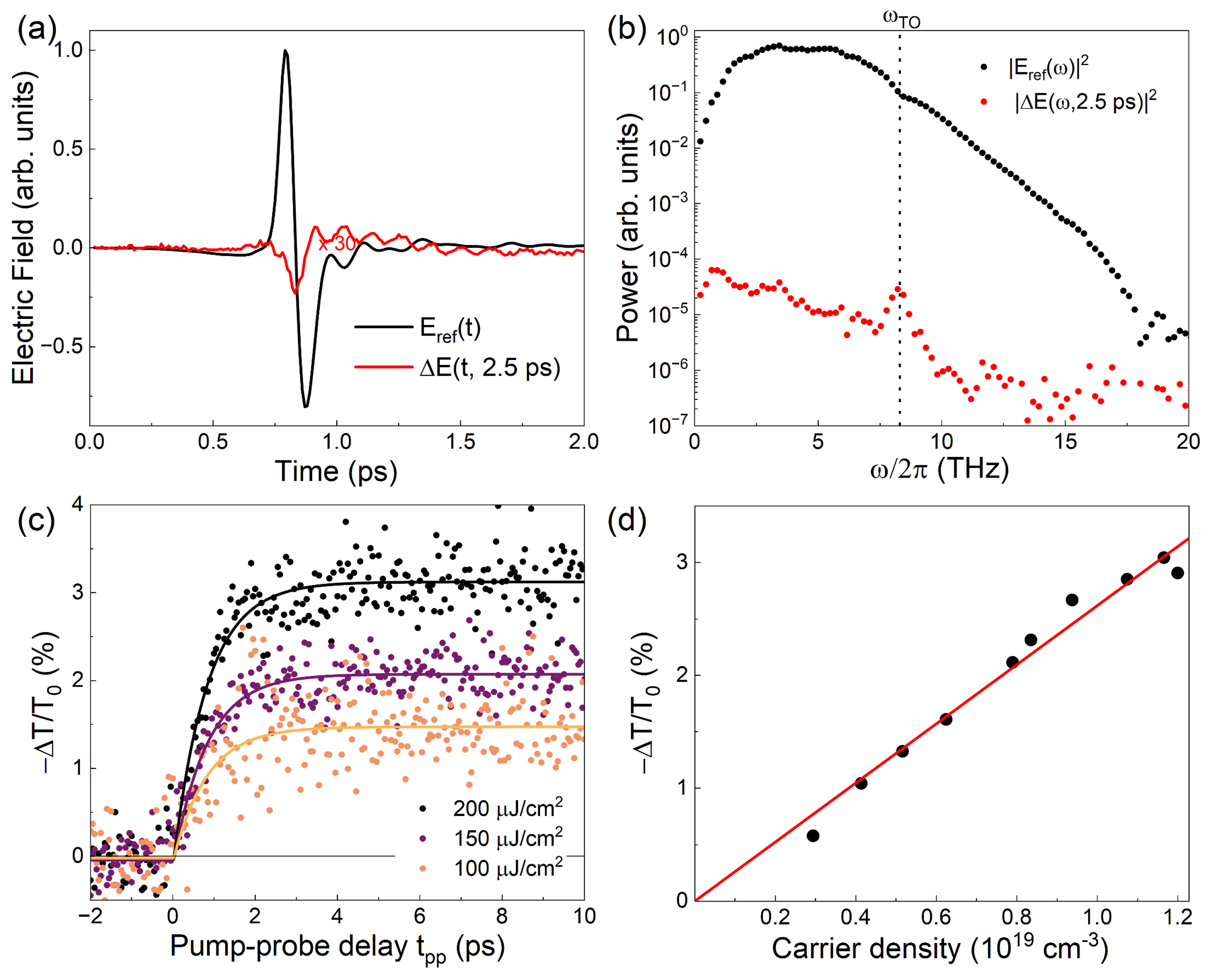}
    \caption{(a) Reference $E_\mathrm{ref}$ and differential $\Delta E$ THz waveforms taken 2.5~ps after excitation. (b) The Fourier power spectra of these pulses, showing the relatively small modulation and a phonon resonance at 8.3~THz. (c) Time-resolved negative differential transmission dynamics (-$\Delta T/T_0$) measured as a function of pump-probe time delay for different pump fluences. The solid lines are fits to the experimental data (circles) using a Heaviside-switched exponential rise convoluted with a Gaussian instrument response function of 35~fs FWHM. (d) Peak $-\Delta T/T_0$ as a function of carrier density $N$ and the linear fit confirming a one-photon absorption process.}
 \label{fig2}
\end{figure}

\begin{figure*}[t!]
    \includegraphics[width=1.0\textwidth]%
    {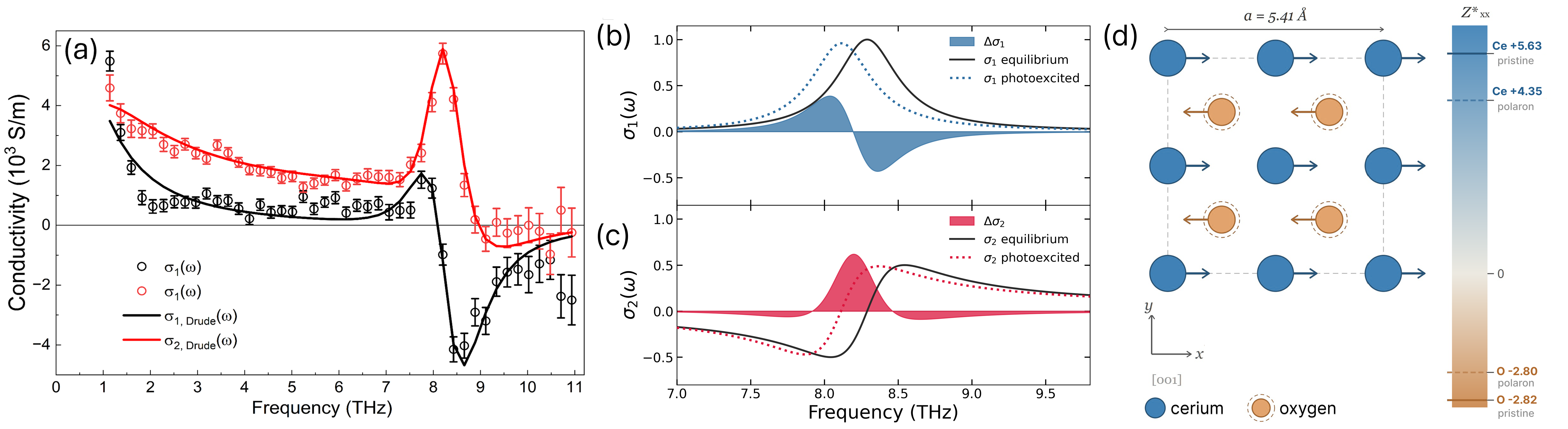}
    \caption{(a)~Complex photoconductivity change 
    $\Delta\sigma(\omega)$ extracted from the THz 
    transmission spectra. Open circles represent the 
    experimental data with error bars, while the solid 
    curves show a combined Drude--Lorentz model fit.
    Real (b) and imaginary (c) part of the conductivity in the phonon region (7--9.5~THz), calculated from Eqs.~(3) and (4) using the equilibrium and photoexcited Drude--Lorentz parameters extracted from the fit in (a) (Table~I).
   % : equilibrium $\sigma_1^{\rm eq}(\omega)$ (black solid) and photoexcited $\sigma_1^{\rm pe}(\omega)$ (blue dotted); shaded area shows $\Delta\sigma_1(\omega)$. 
   The phonon shift is exaggerated for visual clarity. 
    %the true photo-induced change corresponds to the fitted values in Table~I ($\Delta S/S_0 \sim -2$ to $-3\%$). 
    %(c) Corresponding imaginary part $\sigma_2(\omega)$, computed in the same way from Eqs.~(3) and (4), with the same exaggerated scaling as in (b); shaded area shows $\Delta\sigma_2(\omega)$.
    (d) Schematic of the $F_{1u}$ infrared-active 
    phonon mode of CeO$_2$ in the [001] projection of the fluorite structure ($Fm\bar{3}m$, $a = 5.41$~\AA), with Ce atoms displaced along $+x$ and O atoms displaced along $-x$. The colour bar encodes the diagonal Born effective 
    charge $Z^*_{xx}$ obtained from \textit{ab initio} DFPT calculations: solid markers display the pristine-cell 
    values ($Z^*_\mathrm{Ce} = +5.63\,e$, 
    $Z^*_\mathrm{O} = -2.82\,e$) and dashed markers 
    denote values computed for the Ce$^{3+}$ polaron 
    supercell ($Z^*_\mathrm{Ce^{3+}} = +4.35\,e$, 
    $Z^*_\mathrm{O,nn} = -2.80\,e$), showing how 
    polaron formation reduces the dynamical charge of 
    Ce and thereby suppresses the oscillator strength 
    of the $F_{1u}$ mode. }
    \label{fig:fig3}
\end{figure*}

In wide band gap oxides, photoexcited charge carriers can couple strongly to LO phonons to form polarons that can 
significantly alter transport and catalytic properties~\cite{FranchiniNatRevMat2021}. \ce{CeO2} possesses one of the largest static dielectric constants ($\varepsilon_s = 25$) among the rare-earth oxides~\cite{Fromhold1976}, and the highly polar nature of the lattice is reflected in a large splitting between its transverse-optical (TO)
($\hbar\omega_\mathrm{TO} = 34$~meV) and longitudinal-optical (LO)
($\hbar\omega_\mathrm{LO} = 74$~meV) phonon modes~\cite{Weber1993,Mochizuki1982,Marabelli1987}.
%Indeed, the polar nature  of \ce{CeO2} is captured directly by the charge-transfer physics present in its electronic structure \cite{https://doi.org/10.1103/PhysRevLett.55.418}, with the conduction band heavily dominated by \ce{Ce} 4$f$-orbitals and the valence band by \ce{O} 2$p$-orbitals (see Fig.~\ref{fig:fig1}b).
The strength of electron ($e$) and hole ($h$) phonon interactions in 
polar media, which in turn impacts on their band properties and mobility, is characterized by the dimensionless Fr\"{o}hlich coupling constant
\begin{equation}
    \alpha_{e,h} = \sqrt{\frac{m^*_{e,h}}
    {2\hbar\omega_\mathrm{LO}}}
    \frac{e^2}{4\pi\hbar\varepsilon_0}
    \biggl(\frac{1}{\varepsilon_\infty} 
    - \frac{1}{\varepsilon_s}\biggr),
    \label{eq:alpha_def}
\end{equation}
where $\varepsilon_\infty = 4.7$--$7$ is the estimated high-frequency dielectric constant across experimental and theoretical studies of \ce{CeO2} \cite{Marabelli1987,Mochizuki1982,Wang2013, Yamamoto2005}; throughout this work we adopt 
$\varepsilon_\infty = 5.3$~\cite{Mochizuki1982}. A marked asymmetry between the two carrier species then becomes evident: electrons in the flat Ce~$4f$ conduction band ($m^*_e \approx 5$--$10\,m_e$, $\alpha_e \approx 4.5$--$6.4$) form localized Holstein-type small polarons, while holes in the dispersive O~$2p$ valence band ($m^*_h \approx 1.5$--$1.9\,m_e$, 
$\alpha_h \approx 2.5$--$2.9$) are predicted to remain delocalized as Fr\"{o}hlich-type large polarons~\cite{Jiang2025}. Other studies have shown hole localization in p-type \ce{CeO2} \cite{Keating2012}.
\\
\indent The dual (both electron and hole) polaronic nature of photoexcited \ce{CeO2} remains to be experimentally confirmed. Only recently has work begun to elucidate the ultrafast dynamics of the electron polaron in \ce{CeO2} with binding energies of $\sim\!400$~meV and sub-picosecond formation times following optical injection into the Ce 4f band~\cite{Pelli2020, Katoch2024}. Conversely, the nature of the hole polaron remains experimentally unknown although recent calculations predict formation energies on the order of 40~meV \cite{Jiang2025}. Transient optical spectroscopies are sensitive to the localized Ce$^{3+}$ electron states, but they do not directly measure the low-energy intra-band response of mobile carriers. 
%Moreover, the hole mobility for a large polaron at intermediate coupling ($\alpha_h \sim 2$--$3$) is itself an open question, with the Kadanoff Boltzmann equation~\cite{Kadanoff1963} and path-integral formulation~\cite{Feynman1962}, recently applied extensively to lead halide perovskites~\cite{Frost2017,Bretschneider2018}, predicting mobilities of the order of 10--20~cm$^2$\,V$^{-1}$\,s$^{-1}$ for the coupling strengths and phonon energies relevant to \ch{CeO2}, well below the bare Fr\"{o}hlich scattering limit of $\sim\!57$~cm$^2$\,V$^{-1}$\,s$^{-1}$ (details concerning these calculated values can be found in the Supplemental Material). 
In this work, we use ultrabroadband time-resolved terahertz spectroscopy (TRTS)~\cite{Jepsen2011} to probe the photoinduced carrier dynamics in a 50~nm \ce{CeO2} thin film following femtosecond UV excitation. We simultaneously resolve the delocalized Drude response of the hole and directly observe the electron-induced renormalization of the $F_{1u}$ TO phonon, allowing the contributions of mobile holes and self-localized electrons to be disentangled.\\
A 50~nm-thick \ch{CeO_2} thin film sample was grown by reactive magnetron sputtering on a (100)-oriented single crystal diamond substrate (of $\sim$0.5~mm thickness, 3~mm $\times$ 3~mm in size). The film was polycrystalline in nature with nearly stoichiometric composition. Further details on sample growth and characterization are provided in the Supplemental Material.\\
%Cerium oxide was evaporated by a \ch{CeO_2} target in a flux of 20 sccm of an \ch{O_2}/Ar mixture with 75\% \ch{O_2}, applying a power of 150 W, maintaining the sample at room temperature during the growth process to optimize the film adhesion with the diamond substrate \KHB{(KHB: Don't quite follow the notation highlighted in red in this sentence, typo?)}. The film thickness was determined using a quartz microbalance, calibrated by measuring the actual thickness of a cerium oxide film with a profilometer. To permit preliminary static characterization, simultaneously with the diamond substrate, \ch{CeO_2} was evaporated on a Si substrate with thermal oxide for X-ray photoemission spectroscopy and X-ray diffraction. For UV–vis spectrophotometry a UV-grade fused silica substrate was instead chosen, for its optical transparency in the measured window. The measurement details, shown in the Supplemental Material \cite{SupplementalMaterial}, revealed that the sample is polycrystalline, nearly stoichiometric and with an optical absorption spectrum compatible with the values observed on similar \ch{CeO_2} films \cite{Pelli2020, Pelatti2025}.
 \indent TRTS measurements were performed using an amplified Ti:sapphire femtosecond laser providing 5~mJ, 35~fs duration pulses with center wavelength of 795~nm at 1~kHz repetition rate. A two-color laser air-plasma THz source generated single-cycle, broadband THz pulses \cite{KimOptExp2007}, shown in Fig.~\ref{fig2}(a) with the power spectrum shown in Fig.~\ref{fig2}(b), and these were coherently detected using an air-biased coherent detection (ABCD) scheme~\cite{Dai2006}. The third harmonic of the laser amplifier output, producing 267~nm (4.64~eV) ultraviolet pump pulses, was used to photoexcite at a fluence of up to 250~$\mu$J/cm$^2$ collinearly with the THz probe beam. At this photon energy, the excitation promotes electrons from the O~$2p$ valence band into the Ce~$4f$ conduction band (see Fig.~\ref{fig:fig1}(b)), creating electron–hole pairs whose subsequent dynamics are probed by the THz pulse. All measurements were performed at room temperature (295~K) and under dry-air conditions. A schematic of the experimental setup is shown in the Supplemental Material.\\
%THz pulses were generated by co-focusing 800~nm, 35~fs pulses from a regenerative Ti:sapphire amplifier and their second harmonic into dry air. The generated THz pulses were collected and focused by a pair of parabolic mirrors onto the \ch{CeO2} sample, and the transmitted pulses were collimated and focused by a second pair of parabolic mirrors onto the detector. The transmitted THz transients were detected using an air-biased coherent detection (ABCD) scheme, in which an 800~nm gate pulse is focused collinearly with the THz pulse, and the second-harmonic signal generated in the ac bias field serves as a phase-sensitive probe of the instantaneous THz electric field~\cite{Dai2006}. 
%The UV pump excitation spot had a slightly elliptical Gaussian profile with $1/e^2$ radii of $w_x = 1.4$~mm and $w_y = 1.0$~mm and a pump fluence of approximately 200~$\mu$J/cm$^2$.
\indent Photoconductivity dynamics were measured using one-dimensional (1D) differential THz transmission where only the peak of the transmitted THz pulse electric field is monitored as a function of pump-probe delay time $t_{pp}$. The negative differential transmission $-\Delta T(t_\mathrm{pp})/T_0 = 
-\bigl(E_{\mathrm{pump}}(t_\mathrm{pp})-E_{\mathrm{ref}}\bigr)/
E_{\mathrm{ref}}$ is a measure of the averaged THz response over the bandwidth of the pulse and captures the dynamics well. The pump-induced differential THz transmission is shown in Fig.~\ref{fig2}(c), $-\Delta T/T_0$, as a function of pump-probe delay for pump fluences of 100, 150 and 200~$\mu$J/cm$^2$. Fig.~\ref{fig2}(d) confirms the linearity of the peak $-\Delta T/T_0$ response, consistent with the assumption that the photoconductivity arises solely due to photocarriers in the \ch{CeO2} film and not two-photon absorption in the diamond substrate. For all fluences, the differential response is well described by a Heaviside-switched exponential rise convoluted with a 35~fs FWHM Gaussian instrument response. The rise times were found to be independent of fluence with a time constant of $820\pm50$~fs, consistent with self-trapping and not diffusion limited kinetics. We assign the risetime to the formation time of the hole polaron, to be compared to reported bare electron polaron formation times of $~\sim330$~fs \cite{Pelli2020}. Since polaron formation times are in part determined by how quickly the lattice can reorganize around the carrier to build the confining potential well, which depends on the strength of the carrier-phonon coupling, it is reasonable that the hole polaron formation is longer than the electron polaron by a factor of $\alpha_e/\alpha_h\approx2.5$.\\
Full frequency-resolved THz spectroscopy was performed at $t_{pp}=2.5$~ps after photoexcitation, avoiding intrinsic artifacts near $t_{pp}=0$ when phonon lineshapes are dynamically modified \cite{DringoliArXiv2025}. In a frequency-resolved TRTS measurement, the full transmitted THz electric-field in the presence of the pump pulse $E_{\mathrm{pump}}(t,t_{pp})$ and the differential waveform $\Delta E(t,t_{pp})= E_{\mathrm{pump}}(t,t_{pp})-E_{\mathrm{ref}}(t)$ are measured simultaneously through a double modulation technique, where $E_{\mathrm{ref}}(t)$ is the reference waveform in the absence of excitation. An example of the $E_\mathrm{ref}$ and $\Delta E$ waveforms is shown in Fig.~\ref{fig2}(a), displaying an inductive lag by $\approx 90^\circ$ relative to the reference pulse and an $8.3$~THz ringing due to coupling to a single TO phonon, also shown in the power spectrum shown in the inset. Fourier transformation of the THz time-domain waveforms shown in Fig.~\ref{fig2}(a) yields the reference $\tilde{E}_{\mathrm{ref}}(\omega)$ and pumped $\tilde{E}_{\mathrm{pump}}(\omega,t_{pp})$ frequency-domain spectra. The complex transmission coefficient is related analytically to the transient conductivity of the excited layer by \cite{Tinkham1956}
\begin{equation}
    \tilde{T}(\omega) = \frac{\tilde{E}_{\mathrm{pump}}}{\tilde{E}_{\mathrm{ref}}}= \frac{1 + n_{\mathrm{sub}} +  Z_0 d\ \tilde{\sigma}_{ref}(\omega) }{1 + n_{\mathrm{sub}} + Z_0 d\,  \tilde{\sigma}_{\mathrm{pump}}(\omega,t_{pp})},
 \label{eq:transmission}
\end{equation}
where $n_{\mathrm{sub}}$ is the refractive index of the diamond substrate in the THz range ($n_{\mathrm{sub}}= 2.38$), $Z_0 = 377~\Omega$ is the impedance of free space, and $d$ is the effective photoexcited layer thickness, approximated by the optical absorption depth of \ch{CeO_2} at 267~nm ($\sim$27~nm, see the Supplemental Material. \\
\indent The transient differential complex conductivity spectrum $\Delta\tilde{\sigma}(\omega,t_{pp})$ is shown in  Fig.~\ref{fig:fig3}(a). The spectra are comprised of two components, a free-carrier Drude response and localized oscillator representing the F$_{1u}$ phonon. We model the complex transmission coefficient in Eq.~(\ref{eq:transmission}) assuming the
unpumped conductivity $\tilde{\sigma}_{\mathrm{ref}}(\omega)$ arises solely from the triply degenerate infrared-active $F_{1u}$ 
phonon mode at $\omega_0/2\pi = 8.3\,\mathrm{THz}$, in which O atoms vibrate 
against the heavier Ce atoms, described by~\cite{Marabelli1987}
\begin{equation}
\tilde{\sigma}_{\mathrm{ref}}(\omega)
= \frac{-i\epsilon_0 S_0 \omega_0^2 \omega}
{\omega_0^2 - \omega^2 - i\omega\Gamma_0}
 \label{eq:sigma_ref}
\end{equation}
% exp(-iwt) convention
% Oscillator strength S_0 is dimensionless
with oscillator strength $S_0$, resonance frequency $\omega_0$, and damping rate $\Gamma_0$. Upon UV photoexcitation, the total pump conductivity is the sum of a free-carrier (Drude) contribution (first term, $\tilde{\sigma}_{\mathrm{e}}$) and a modified phonon background (second term, $\tilde{\sigma}_{\mathrm{ph}}$) in the following equation:
\begin{equation}
  \tilde{\sigma}_{\mathrm{pump}}(\omega)
  = \frac{\omega_p^2 \varepsilon_0}{\gamma - i\omega} +  
  \frac{-i S_1 \omega_1^2 \omega \varepsilon_0 }
{\omega_1^2 - \omega^2 - i\omega\Gamma_1},
  \label{eq:drude_lorentz}
\end{equation}
with $\omega_p$ being the plasma frequency, $\gamma = 1/\tau$ is the momentum scattering rate, and $S_1,\,\omega_1$ and $\Gamma_1$ are the phonon parameters under excitation. The differential conductivity probed by the THz field is then $\Delta\tilde{\sigma}(\omega)
  = \tilde{\sigma}_{\mathrm{e}}(\omega)
  + \bigl[\tilde{\sigma}_{\mathrm{ph}}(\omega)
          - \tilde{\sigma}_{\mathrm{ref}}(\omega)\bigr]$
where the bracket represents the photo-induced change in the phonon contribution. The phonon parameters $(S_0,\,\omega_0,\,\Gamma_0)$ are fixed to the values measured by infrared reflectivity on stoichiometric \ch{CeO2} single crystals~\cite{Weber1993}, while the five parameters $(S_1,\,\omega_1,\,\Gamma_1,\,\omega_p,\,\tau)$ are optimised against the measured complex-valued $\tilde{T}(\omega)$ using a nonlinear least-squares minimization weighted by the experimental error bars.

Drude fits yielded a hole momentum 
scattering time $\tau \approx 110$--$160$~fs and a 
plasma frequency 
$\omega_p/2\pi \approx 12$--$14$~THz.
The extracted Drude parameters place the system in the 
inductive transport regime $\omega\tau \gg 1$ over the 
probed THz band. Lorentzian fits to the $F_{1u}$ phonon reveal small but consistent photo-induced shifts in the phonon background: the oscillator strength decreases by $\Delta S/S_0 \sim -2$ to $-3\%$ and the resonance softens by 0.02--0.08~THz ($< 1\%$ of $\omega_0$). The full parameter table is given in Table~\ref{tab:drude_params}. 
\\
\indent These observations raise two immediate questions: which carrier species drives the Drude response, and what is the physical origin of the phonon shifts? To answer the first question, we turn to the electronic structure of \ch{CeO2}. 
The Ce~$4f$ conduction band of \ch{CeO2} is known to be exceptionally flat, with a localized Ce~$4f$ conduction band, while the O~$2p$ valence band is dispersive \cite{DaSilva2007,Skorodumova2001}.
Our \textit{ab initio} PBE0 hybrid DFT calculations 
(see Fig.~\ref{fig:fig1}(b)) confirm this picture, yielding $m^*_e \approx 5.3\,m_0$ and $m^*_h \approx 1.7\,m_0$ from a Kane non-parabolic dispersion fit at the valence band maximum (see the Supplemental Material for more details). As discussed in the introduction, the corresponding 
Fr\"{o}hlich coupling constants place the holes in 
the intermediate-coupling regime 
($\alpha_h \approx 2.5$--$2.9$), where they are 
expected to form mobile large polarons with a 
well-defined Drude 
conductivity~\cite{Devreese2003}, and the electrons in 
the strong-coupling regime 
($\alpha_e \approx 4.5$--$6.4$), where they rapidly 
self-localize as Ce$^{3+}$ small polarons on 
sub-picosecond timescales~\cite{Pelli2020,Katoch2024}.
We therefore attribute the Drude component to the photoinjected holes. The corresponding hole mobility $\mu_h=e\tau/m_{pol,h}^*\approx100$~cm$^2/$Vs assuming a dressed polaron mass $m_{pol,h}^*\approx2.7m_0$ (See Supplemental for calculations). The bare LO-phonon-limited value of $\mu_\mathrm{bare} \approx 57$~cm$^2$\,V$^{-1}$\, s$^{-1}$, the Kadanoff Boltzmann equation result of $\mu_K \approx 20$~cm$^2$\,V$^{-1}$\,s$^{-1}$~\cite{Kadanoff1963}, and the FHIP path-integral mobility of $\mu_\mathrm{FHIP} \approx 
10$~cm$^2$\,V$^{-1}$\,s$^{-1}$~\cite{Feynman1962} (see Supplemental Material for details). The systematic underestimation by all three models, which represent progressively more sophisticated treatments of the polaron--phonon interaction, points to a breakdown of the single-mode Fr\"{o}hlich Hamiltonian at intermediate coupling~\cite{Devreese2007,Mishchenko2019}. Electrons, by contrast, are self-trapped as Ce$^{3+}$ small polarons with mobilities orders of magnitude lower, governed by thermally activated hopping rather than band-like transport~\cite{FranchiniNatRevMat2021}.\\ 
The carrier density $n=\omega_p^2\epsilon_0m_{pol,h}^*/e^2\approx5.7\times10^{18}$~cm$^{-3}$ is to be compared to the absorbed photon flux $N_{max}=F\lambda_p/hcd\approx4.0\times10^{19}$~cm$^{-3}$, yielding a mobile carrier fraction $\eta = n/N_{max}\approx14\%$. This low quantum efficiency can be explained by either the creation of neutral excitons immediately after excitation, or the presence of traps related to grain boundaries and defects within the film. Evaluating the Saha equation at room temperature with an effective screening $\varepsilon_\mathrm{eff} = \sqrt{\varepsilon_\infty\,\varepsilon_s} \approx 11.5$~\cite{Haken1959} gives a free-carrier fraction $\xi \approx 23\%$, in reasonable agreement with the measured $\eta \approx 14\%$ 
(see Supplementary Material for details). Spatially resolved micro-Raman maps also confirm the presence of defects in the photoexcited film~(Fig. S3 in the Supplemental Material), which are absent in a reference unexcited film grown under similar conditions. Cumulative photoreduction drives permanent structural changes~\cite{Wu2019SciRep,Shcherbakov2018} in such a photocatalytic system, as evidenced by the appearance of defect-activated Raman bands 
between 530 and 600~cm$^{-1}$ (Supplemental Material, Fig.~\ref{fig:staticCharacterization}). Indeed, such degradation phenomena are frequently observed in oxides exposed to sustained UV radiation \cite{Kayani2021}.\\
%The time-domain signature reinforces this picture: in the inductive limit, $\Delta E(t) \propto \int_{-\infty}^{t} E_\mathrm{ref}(t')\,dt'$, so the differential field resembles the running time integral of the reference pulse, which is consistent with the $\approx 90^\circ$ lag noted above and precisely what is observed in Fig.~\ref{fig2}a.\\
\begin{table}[b!]
\centering
\caption{%
   Drude and phonon parameters extracted from constrained
  Drude--Lorentz fits to the THz transmission at
  $\tau_{\mathrm{pp}} = 2.5\,\mathrm{ps}$ for two pump--fluence
  conditions.
  The reference phonon parameters are fixed from
  Ref.~\cite{Weber1993}. Uncertainties represent one standard deviation (\(1\sigma\)) from the covariance matrix of the least-squares fit.
}
\label{tab:drude_params}
\begin{tabular}{lcc}
\hline\hline
Parameter & Run 1 & Run 2 \\
\hline
Pump fluence ($\mu$J\, cm$^{-2}$) & 150 & 250  \\
\hline
\multicolumn{3}{l}{\textit{Drude parameters}} \\
$\omega_p/2\pi$ (THz)            & $11.35 \pm 0.17$ & $13.70 \pm 0.19$ \\
$\tau$ (fs)                       & $157 \pm 12$   & $117 \pm 7$   \\
$\chi^2/\mathrm{dof}$            & $1.44$  & $1.35$  \\
\hline
\multicolumn{3}{l}{\textit{Phonon (pump) - shifts from reference}} \\
$\Delta S / S_0$ (\%)            & $-2.4 \pm 0.8$  & $-2.0 \pm 1.8$  \\
$\Delta\omega_0/2\pi$ (THz)      & $-0.070 \pm 0.008$ & $-0.021 \pm 0.039$ \\
$\Delta\Gamma/2\pi$ (THz)        & $+0.028 \pm 0.019$ & $+0.249 \pm 0.064$ \\
\hline
\multicolumn{3}{l}{\textit{Reference phonon }} \\
$S_0$                            & \multicolumn{2}{c}{$14.91$} \\
$\omega_0/2\pi$ (THz)            & \multicolumn{2}{c}{$8.29$}  \\
$\Gamma_0/2\pi$ (THz)            & \multicolumn{2}{c}{$1.40$}  \\
$\varepsilon_\infty$ & \multicolumn{2}{c}{$5.3$~\cite{Mochizuki1982}}  \\
\hline\hline
\end{tabular}
\end{table}

Electrons injected into the flat Ce~$4f$ band 
self-localize rapidly \cite{Pelli2020, Katoch2024}; their intrinsic spectral weight 
shifts to the mid-infrared, above the upper limit of 
our probe frequency. 
Electron--phonon coupling can directly 
modify the phonon properties: the photoinjection of 
Ce$^{4+} \to$ Ce$^{3+}$ electron polaron sites, each 
distorting its local oxygen environment, modifies the vibrational characteristics of the lattice.
The reduction in oscillator strength $S_0$ is understood through the Born effective charge (BEC) of the infrared-active $F_{1u}$ mode: the oscillator strength 
scales as the square of the mode-effective dynamical 
charge (see Supplemental Material for details), 
and any reduction in the BEC at the polaron site 
directly suppresses $S_0$.
Figure~\ref{fig:fig3}(c) illustrates the connection.
In this mode, the Ce and O sublattices undergo opposing 
displacements, generating a macroscopic ionic polarisation 
whose magnitude is governed by the Born effective charge 
$Z^*$~\cite{Gonze1997,Baroni2001}.
The pristine cell value 
$Z^*_\mathrm{Ce} = +5.63\,e$ substantially exceeds 
the formal ionic charge of $+4e$, reflecting the 
dynamic Ce--O charge transfer during the $F_{1u}$ 
displacement: as Ce moves, the O~$2p$--Ce~$4f/5d$ 
hybridization changes, transferring additional electron 
density in the direction of the displacement and 
amplifying the effective dipole. Upon Ce$^{4+} \to$ Ce$^{3+}$ polaron formation, the additional $4f$ electron partially screens the Ce nucleus and partially satisfies the Ce--O bonding demand, reducing the covalent charge transfer that occurs during an ionic displacement and bringing $Z^*(\mathrm{Ce})$ closer to the nominal ionic value.

DFPT calculations on the polaron 
supercell (see the Supplemental Material for details) confirm this and show that upon 
Ce$^{4+} \to$ Ce$^{3+}$ conversion, the isotropic BEC 
at the polaron site decreases from 
$Z^*_\mathrm{Ce} = +5.63\,e$ to 
$Z^*_\mathrm{Ce^{3+}} = +4.35\,e$, a reduction of 
$\Delta Z^*/Z^* \approx -23\%$, bringing it 
significantly closer to the formal ionic value of 
$+3e$.
This reduction reflects a weakening of the dynamic 
Ce--O covalency as the localized $4f$ electron 
partially satisfies the bonding demand and reduces the 
$p$--$f$ charge transfer during an ionic displacement.
%\color{teal}
Projecting the site-resolved BEC changes onto the 
$F_{1u}$ phonon eigenvector gives a mode-averaged 
oscillator strength reduction of 
$\Delta S/S \approx -4\%$ at the supercell polaron 
concentration (1/32), which is consistent in sign and order of magnitude with the experimentally observed 
$-2\%$.
%A direct numerical comparison must be made with care, as the DFPT calculation represents a static equilibrium configuration with one polaron per 32 cerium sites, whereas the experiment probes a transient, non-equilibrium photoexcited state at a different polaron fraction.
\color{black}
The resonance softening 
$\Delta\omega_0 < 0$ has the same 
microscopic origin: at the photoexcited carrier 
density of 
$n \approx 5.7 \times 10^{18}$~cm$^{-3}$, 
approximately one in $10^{4}$ Ce sites captures an 
electron and converts to Ce$^{3+}$.
At each such site, the surrounding oxygen cage 
expands due to the larger ionic radius of Ce$^{3+}$ 
($1.14$~\AA\ vs $0.97$~\AA\ for 
Ce$^{4+}$~\cite{Shannon1969}), reducing the local 
Ce--O bond stiffness and softening the restoring 
force acting on the O sublattice.

\indent In summary, we have used ultrabroadband time-resolved THz spectroscopy to directly probe photoinduced polarons in \ch{CeO2} following above-bandgap excitation at $\lambda_p=267$~nm. Transient spectra reveal Drude-like conduction of Fr\"ohlich-type hole-polarons with mobilities of 100~cm$^2$/Vs, while small electron polarons are revealed through the induced modification to the lattice Born effective charge and renormalization of the phonon frequency and linewidth. This simultaneous observation of both polaron species within a single broadband measurement constitutes a direct experimental measurement of both small and large polarons within a single photocatalytic material~\cite{Jiang2025}. This work paves the way for future TRTS studies of charge transfer oxides, probing intrinsic charge transport and coupling to the lattice. 

\section*{Acknowledgments}
%\begin{acknowledgments}

VM and DGC gratefully acknowledge funding from NSERC, FRQNT and CFI. Computational resources were provided by the Digital Research Alliance of
Canada. VM and DGC also thank Peter Elliott (Science and Technology Facilities Council, UK) for helpful discussions regarding the DFT band structure calculations. 

%\end{acknowledgments}
% Add the supplemental here ONLY for arxiv submission
\section*{Supplemental Material}
\subsection{Time-resolved THz spectrometer}
\begin{figure}
    \centering
\includegraphics[width=0.9\columnwidth]{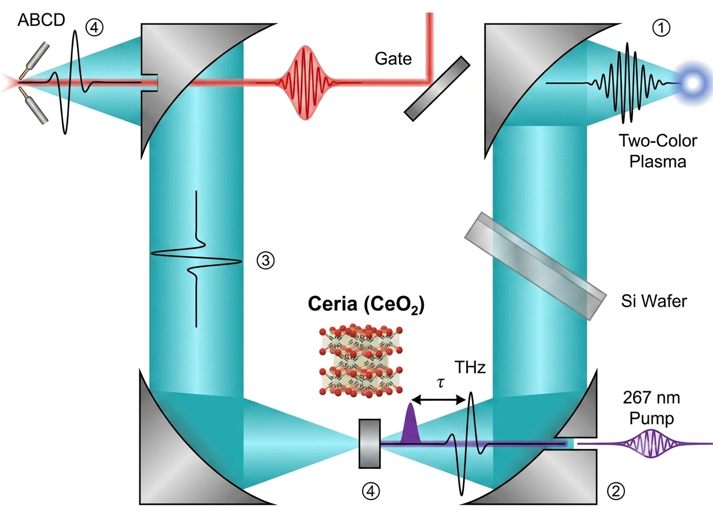}
    \caption{Schematic of the ultrabroadband time-resolved 
    THz spectroscopy (TRTS) setup.}
    \label{sm:setup}
\end{figure}

The experimental setup for time-resolved terahertz (THz) spectroscopy is shown in Fig.~\ref{sm:setup}. THz pulses were generated by co-focusing 800~nm, 35~fs pulses from a regenerative Ti:sapphire amplifier and their second harmonic into dry air. The generated THz pulses were collected and focused by a pair of parabolic mirrors onto the \ch{CeO2} sample, and the transmitted pulses were collimated and focused by a second pair of parabolic mirrors onto the detector. The transmitted THz transients were detected using an air-biased coherent detection (ABCD) scheme, in which an 800~nm gate pulse is focused collinearly with the THz pulse, and the second-harmonic signal generated in the ac bias field serves as a phase-sensitive probe of the instantaneous THz electric field~\cite{Dai2006}. 
%The UV pump excitation spot had a slightly elliptical Gaussian profile with $1/e^2$ radii of $w_x = 1.4$~mm and $w_y = 1.0$~mm and a pump fluence of approximately 200~$\mu$J/cm$^2$

\subsection{Static characterization}
\label{appendix:static_characterization}
X-ray photoelectron spectroscopy (XPS) and X-ray diffraction (XRD) characterizations were performed on a \ch{CeO_2} film deposited on a Si substrate with thermal oxide, while ultraviolet-visible spectrophotometry measurements were done on a \ch{CeO_2} film deposited on a UV-grade fused silica substrate, to guarantee optical transparency. The two substrates were cleaned by immersing them in a bath of boiling acetone for 5 min, followed by an ultrasonic bath in boiling acetone and by an ultrasonic bath in boiling isopropanol for 3 min each. All samples were grown simultaneously with the sample on the diamond substrate used for the dynamic characterization. 

\begin{figure}
    \centering
    \includegraphics[width=0.9\columnwidth]{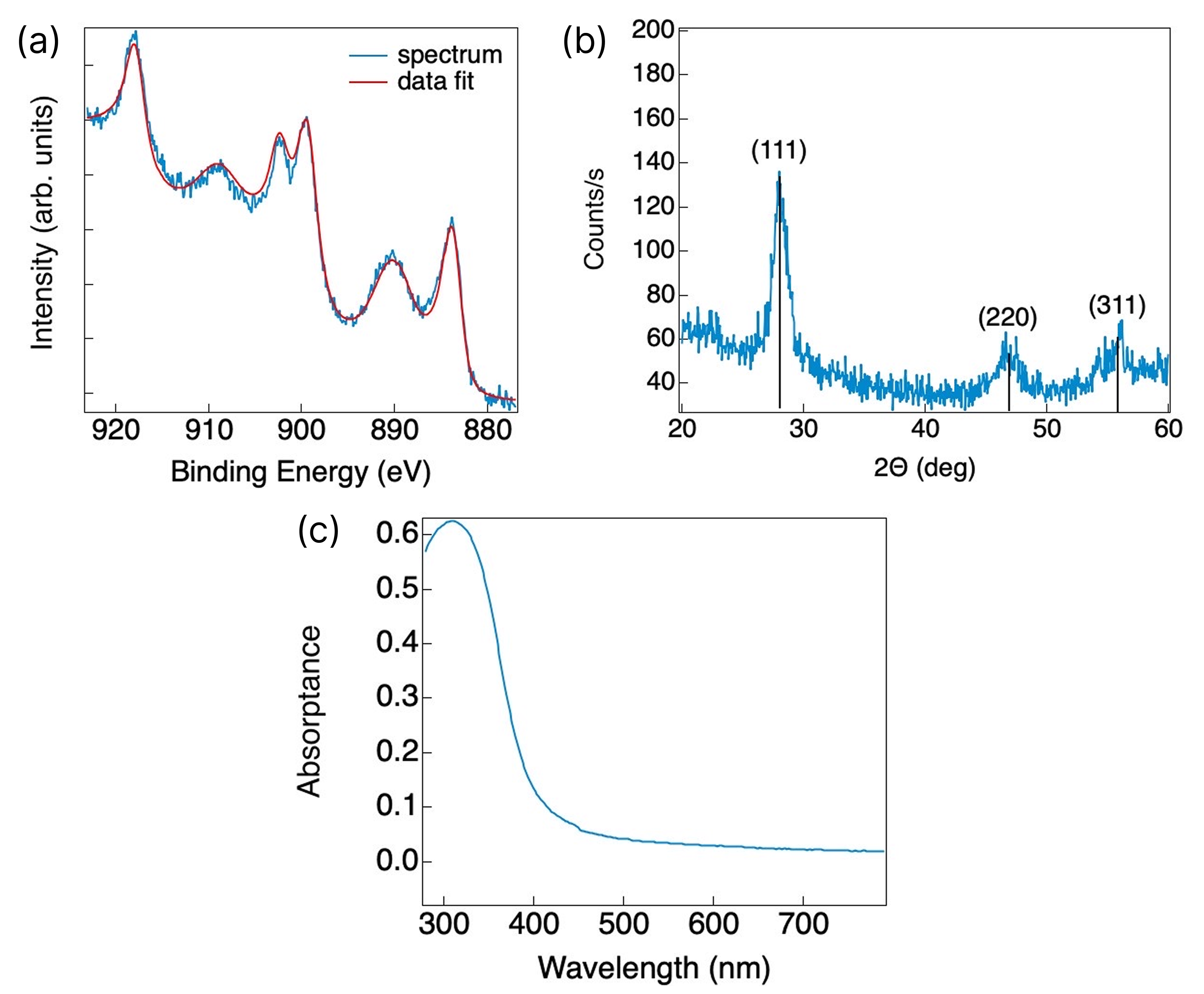}
    \caption{Static characterization of the \ce{CeO2} thin film (a) XPS Ce~$3d$ emission spectrum 
             (b) XRD characterization
             (c) UV-vis optical absorptance.}
    \label{fig:staticCharacterization}
\end{figure}

Figure~\ref{fig:staticCharacterization}a reports the XPS Ce 3d emission line. The XPS spectrum was acquired using an Al K$_{\alpha}$ photon source and a hemispherical electron analyzer. To obtain information on the surface stoichiometry of the cerium oxide film, the Ce 3d spectrum was fitted following reference \cite{Skla2008}, obtaining a dominant \ch{CeO_2} stoichiometry with a \ch{Ce^{3+}} concentration below the detection limit. The XRD profile of the \ch{CeO_2} film reported in Fig. \ref{fig:staticCharacterization}b shows a dominant peak at $\simeq 28^{\circ}$, related to (111) planes, and two minor peaks at $\simeq 47^{\circ}$ and $\simeq 56^{\circ}$ related, respectively, to the (220) and (311) planes (POW COD open diffraction database, card 00-721-7887), suggesting that the film is polycrystalline with a dominant (111) texture. The setup for UV-Vis spectrophotometry includes a Xe lamp, a monochromator, and a Si photodetector. The sample was measured at an incidence angle of $22^{\circ}$ from the surface normal. The optical absorptance shown in Fig.\ref{fig:staticCharacterization}c was calculated as $A = 1 - T - R $, where T and R are the transmittance and reflectance of the cerium oxide film, respectively. 

\subsection{Effective mass of charge carriers in \ch{CeO2}}
\label{appendix:effective_mass}
The effective mass $m^{\ast}$ entering Eq.~\eqref{eq:alpha_def2}
was extracted from the curvature of the calculated electronic band structure along the $\Gamma \rightarrow \mathrm{K}$ high-symmetry path in the Brillouin zone. According to semi-classical transport theory, the inertial effective mass $m_e^*$ of a carrier in a given band is inversely proportional to the second derivative of the energy dispersion $E(k)$ with respect to the wavevector $k$, defined by the relation \( m^*_{e,h} = \hbar^2 \left( \frac{\partial^2 E}{\partial k^2} \right)^{-1}  \), where $\hbar$ is the reduced Planck constant. In practice, deviations from
a purely parabolic dispersion are significant near band extrema in polar
oxides, and a simple parabolic fit overestimates the curvature when applied
over a wide $k$-range. We therefore fit the dispersion to the Kane non-parabolic model \cite{Kane1957}, which is given as
\begin{equation}
  \frac{\hbar^{2}k^{2}}{2m^{\ast}} = E\!\left(1 + \alpha_{\mathrm{np}}\,E\right),
  \label{eq:kane}
\end{equation}
where $\alpha_{\mathrm{np}}$ is the non-parabolicity coefficient
The effective masses were extracted from the PBE0 hybrid band structure using the
\textit{ab initio} scattering and transport (AMSET) package~\cite{ganose2021},
which fits Eq.~\eqref{eq:kane} numerically to the DFT eigenvalues along
a specified $k$-path. The AMSET analysis reveals that the valence band maximum (VBM) of \ch{CeO2}
does not occur at a high-symmetry point, but between $K$ and $\Gamma$ at
$\mathbf{k} \approx (0.27,\,0.27,\,0.53)$ (in reciprocal lattice
coordinates). Similarly,
the conduction band minimum (CBM) lies between $K$ and $\Gamma$ at
$\mathbf{k} \approx (0.34,\,0.34,\,0.67)$.  This gives an indirect band
gap of $E_g^{\mathrm{ind}} = 4.228$~eV, while the smallest direct gap,
located at $\mathbf{k} \approx (0.29,\,0.29,\,0.58)$, is
$E_g^{\mathrm{dir}} = 4.255$~eV, only $27$~meV larger.  The near
degeneracy of the indirect and direct gaps implies that above-bandgap UV
excitation at $4.64$~eV accesses both transitions simultaneously. 

Because the VBM occurs at an off-symmetry point, the band curvature is
inherently asymmetric: the dispersion toward $\Gamma$ and toward $K$ are
independently fit by AMSET along the path
$K \to \mathbf{k}_{\mathrm{VBM}} \to \Gamma$.  The two values are
\begin{equation}
  m_{h,\to\Gamma}^* = -1.931\,m_0, \qquad
  m_{h,\to K}^*     = -1.508\,m_0,
  \label{eq:mh_anisotropic}
\end{equation}
reflecting the asymmetric curvature on either side of the off-symmetry
maximum.  The transport effective mass, which governs carrier mobility via
the Drude relation $\mu_h = e\tau/m_h^*$ and corresponds to the harmonic
mean of the two curvatures, is
\begin{equation}
  m_h^*
  = \left( \frac{1}{2} \cdot \frac{1}{m_{h,\to\Gamma}^*}
          + \frac{1}{2} \cdot \frac{1}{m_{h,\to K}^*}   \right)^{-1}
  \approx -1.7\,m_0.
  \label{eq:mh_transport}
\end{equation}
The AMSET fit yields electron effective masses of
\begin{equation}
  m_{e,\to\Gamma}^* = 5.26\,m_0, \qquad
  m_{e,\to K}^*     = 5.27\,m_0,
  \label{eq:me_values}
\end{equation}
giving a transport effective mass for the electron $m_e^* \approx 5.3\,m_0$ used throughout this work. 
This analysis clearly indicates that the holes in
the O~$2p$ valence band of \ch{CeO2} are lighter than the electrons. Instead, 
the lowest conduction band is of predominantly Ce~$4f$ character and is
exceptionally flat across the entire Brillouin zone, consistent with the localized nature of
the $4f$ orbitals and a larger electron mass. 
\subsection{Fröhlich coupling constant}
\label{appendix:froehlich_coupling}

The long-range interaction between charge carriers and longitudinal optical
(LO) phonons in polar crystals is described by the Fröhlich
Hamiltonian~\cite{Devreese2020}.
Its strength is quantified by the dimensionless Fröhlich coupling constant
$\alpha_{e,h}$, defined for a single dispersionless LO phonon mode of angular
frequency $\omega_{\mathrm{LO}}$ and an isotropic band with effective mass
$m^*_{e,h}$ as~\cite{frohlich1954electrons}
\begin{equation}
    \alpha_{e,h} =
    \frac{e^2}{4\pi\varepsilon_0\hbar}
    \left(\frac{1}{\varepsilon_\infty} - \frac{1}{\varepsilon_s}\right)
    \sqrt{\frac{m^*_{e,h}}{2\hbar\omega_{\mathrm{LO}}}},
    \label{eq:alpha_def2}
\end{equation}
where \(\frac{1}{\varepsilon_\infty}-\frac{1}{\varepsilon_s}\) qualitatively manifests the
capacity of the screening due to a Coulomb potential of the
polarized lattice \cite{Jin2022}. 
For cubic \ch{CeO2} we use $\varepsilon_\infty = 5.3$ from the infrared optical measurements of Mochizuki~\cite{Mochizuki1982} on bulk and thin-film \ce{CeO2}, and $\hbar\omega_\text{TO} = 34$~meV and $\hbar\omega_\text{LO} = 74$~meV 
from Weber~\cite{Weber1993}, from which the LST relation gives 
$\varepsilon_s \approx 25$.
Inserting the hole band mass $m_h^* \approx 1.7\,m_0$
into Eq.~\eqref{eq:alpha_def2} yields \(\alpha_{h}  \approx  2.6\) for holes, which places holes in \ch{CeO2} in the intermediate Fröhlich coupling regime ($1 \lesssim \alpha_{e,h} \lesssim 6$), where neither the weak-coupling perturbative expansion nor the strong-coupling Pekar limit strictly apply ~\cite{Devreese2007}.
The corresponding polaron binding energy is obtained by numerically
minimising the Feynman variational ground-state energy
$E_0(v,w)$ at $\alpha_h$, rather than from the weak-coupling estimate
$E_\mathrm{pol} \approx \alpha\hbar\omega_\mathrm{LO}$, which
underestimates the binding by $\sim$5\% at this coupling strength.
This gives $E_0/\hbar\omega_\mathrm{LO} \approx -2.70$, corresponding to
$E_{\mathrm{pol},h} \approx 200$~meV.
For the electron band mass $m_e^* \approx 5.3\,m_0$, the coupling
constant is $\alpha_{e} \approx 4.6$, placing electrons deeper into the
intermediate regime and making them susceptible to polaronic
self-localization on sub-picosecond timescales.
The variational ground-state energy at this coupling is
$E_0/\hbar\omega_\mathrm{LO} \approx -4.96$, giving
$E_{\mathrm{pol},e} \approx 367$~meV, which is nearly twice the hole binding energy and consistent with electrons undergoing deeper self-trapping.
\subsection{Skin depth of \ch{CeO_2}}
\label{appendix:skin_depth}
We estimate the ultraviolet skin depth of \ce{CeO2} from the measured incident and transmitted pump powers. With an incident power of
\(P_{\mathrm{inc}} = 7.0 \pm 0.2~\mathrm{mW}\) 
and a transmitted power of
\(P_{\mathrm{trans}} = 1.1 \pm 0.2~\mathrm{mW}\),
the Beer--Lambert law
\begin{equation}
    I(z) = I_0 \exp(-\gamma z)
\end{equation}
relates the absorption coefficient \(\gamma\) to the intensity (or power) transmission through a film of thickness \(d\) via
\begin{equation}
    T \equiv \frac{I(d)}{I_0} \simeq \frac{P_{\mathrm{trans}}}{P_{\mathrm{inc}}}
    = \exp(-\gamma d).
\end{equation}
Solving for \(\gamma\) gives
\begin{equation}
    \gamma = \frac{1}{d}\ln\!\left(\frac{1}{T}\right)
          = \frac{1}{d}\ln\!\left(\frac{P_{\mathrm{inc}}}{P_{\mathrm{trans}}}\right).
\end{equation}
For a 50-nm thick \ch{CeO_2}, the above formula yields a skin depth equal to 
\begin{equation}
    \delta = \frac{1}{\gamma} \approx (27 \pm  3) ~\mathrm{nm}.
\end{equation}

Thus, within experimental uncertainty, the UV skin depth in  \(50  ~\mathrm{nm}\)-thick \ch{CeO2} film is approximately \(\delta \simeq 27 ~\mathrm{nm}\). This value is fairly similar to the skin depth of  \(\delta \simeq 36 ~\mathrm{nm}\), reported in \cite{Debnath2007} for \(140  ~\mathrm{nm}\)-thick \ch{CeO_2} samples.

\subsection{Drude--Lorentz fit parameters}
\label{appendix:drude_params}
 
Table~\ref{tab:drude_params} lists the parameters extracted from the constrained Drude--Lorentz fits to the complex THz transmission ratio $\tilde{T}(\omega)$ at a pump--probe delay of $2.5\,\mathrm{ps}$. 
The reference transverse optical (TO) phonon parameters ($S_0 = 14.91$, $\omega_{\text{TO}}/2\pi = 8.29\,\mathrm{THz}$, $\Gamma_0/2\pi = 1.40\,\mathrm{THz}$) are fixed from the bulk infrared-active $F_{1u}$ mode values reported by Weber et al.~\cite{Weber1993} and are common to all fits.
The pump phonon parameters $(S_1,\,\omega_1,\,\Gamma_1)$ are allowed
to vary within the bounds described in the main text.

\subsection{Polaron Mobility Estimates and Scattering 
         Analysis}
\label{appendix:mobility}
The measured hole mobility 
$\mu_h \approx 103$~cm$^2$\,V$^{-1}$\,s$^{-1}$ is 
compared below against three progressively more 
sophisticated theoretical estimates.
All models use the material parameters of \ch{CeO2}: 
$\hbar\omega_\mathrm{LO} = 74$~meV~\cite{Weber1993}, 
$\varepsilon_\infty = 5.3$~\cite{Mochizuki1982}, 
$\varepsilon_s \approx 25$ (from the LST relation), 
$m^*_h = 1.7\,m_0$, and the resulting Fr\"{o}hlich 
coupling constant $\alpha_h \approx 2.6$ 
(Eq.~1 of the main text), which places the holes in 
the intermediate-coupling regime.

\subsubsection{Bare Fr\"{o}hlich scattering rate}
\label{sm:froehlich}
As a first baseline, we estimate the LO-phonon-limited 
momentum relaxation time.
To calculate the electron-phonon scattering rate in a highly ionic lattice like \ch{CeO_2}, the dominant mechanism at room temperature is the Fröhlich interaction with longitudinal optical (LO) phonons. To estimate the characteristic momentum relaxation time associated with LO-phonon scattering, we use the polar optical-phonon scattering rate for a three-dimensional, non-degenerate carrier gas within the Fröhlich model, accounting for band non-parabolicity via the Kane
model~\cite{Ridley2013,Kane1957}.
The energy-dependent momentum scattering rate is then expressed as ~\cite{Ridley2013}

\begin{equation}
\label{eq:tau_full}
\begin{split}
\frac{1}{\tau_m^\mathrm{LO}} &= \frac{e^2(2m^*_{e,h})^{1/2}\omega_{LO}}{8\pi \epsilon_0\varepsilon_{\mathrm{p}} \gamma^{1/2}(E_k)} \\
&\quad \times \Biggl[ N_{LO} \left\{ \frac{\mathrm{d}\gamma(E_k)}{\mathrm{d}E_k} \right\}_{E_k+\hbar\omega_0} \frac{\gamma^{1/2}(E_k + \hbar\omega_0)}{\gamma^{1/2}(E_k)} \\
&\quad + \{N_{LO} + 1\} \left\{ \frac{\mathrm{d}\gamma(E_k)}{\mathrm{d}E_k} \right\}_{E_k-\hbar\omega_0} \frac{\gamma^{1/2}(E_k - \hbar\omega_0)}{\gamma^{1/2}(E_k)} \\
&\quad -N_{LO} \left\{ \frac{\mathrm{d}\gamma(E_k)}{\mathrm{d}E_k} \right\}_{E_k+\hbar\omega_0} \left\{ \frac{\gamma(E_k + \hbar\omega_0) - \gamma(E_k)}{\gamma(E_k)} \right\} \\
&\quad \times \coth^{-1} \left\{ \frac{\gamma^{1/2}(E_k + \hbar\omega_0)}{\gamma^{1/2}(E_k)} \right\} \quad + \{N_{LO}+ 1\} \times\\
& \left\{ \frac{\mathrm{d}\gamma(E_k)}{\mathrm{d}E_k} \right\}_{E_k-\hbar\omega_0} \left\{ \frac{\gamma(E_k) - \gamma(E_k - \hbar\omega_0)}{\gamma(E_k)} \right\} \\
&\quad \times \tanh^{-1}  \left\{ \frac{\gamma^{1/2}(E_k - \hbar\omega_0)}{\gamma^{1/2}(E_k)} \right\} \Biggr]
\end{split}
\end{equation}

where deviations from a parabolic dispersion are captured by the function
\begin{equation}
    \gamma(E_k) = E_k\!\left(1 + \alpha_{\mathrm{np}}\,E_k\right),
    \label{eq:gamma_np}
\end{equation}
and the scattering strength is scaled by the effective polaron dielectric
constant $\varepsilon_p = (1/\varepsilon_\infty - 1/\varepsilon_s)^{-1}$. The terms in the above equation proportional to $N_{\mathrm{LO}}$ and $N_{\mathrm{LO}}+1$
correspond to LO-phonon absorption and emission, respectively.
The non-parabolicity coefficient is related to the band gap $E_g$ and the
carrier effective mass by~\cite{Kane1957}
\begin{equation}
    \alpha_{\mathrm{np}}
    \approx \frac{1}{E_g}\left(1 - \frac{m^*_{e,h}}{m_e}\right)^2,
    \label{eq:alpha_np}
\end{equation}
which for holes with $m^*_h \approx 1.7\,m_0$ and $E_g = 4.26$~eV
gives $\alpha_{\mathrm{np},h} \approx 0.63$~eV$^{-1}$.

The bare band mass $m^*_h = 1.7\,m_0$ is dressed by the 
phonon cloud into the polaron mass $m^*_{\mathrm{pol},h}$.
In the Feynman variational 
model~\cite{Feynman1955,Schultz1959}, the mass 
renormalization is given by
\begin{equation}
    \frac{m^*_{\mathrm{pol},h}}{m^*_h}
    = 1 + \frac{\alpha_h}{3\sqrt{\pi}}
      \int_0^\infty d\tau_0\,
      \tau_0^2 e^{-\tau_0}\,
      \bigl[D(\tau_0)\bigr]^{-3/2},
    \label{eq:feynman_mass}
\end{equation}
where $D(\tau_0) = (w^2/v^2)\tau_0 + 
(1 - w^2/v^2)(1 - e^{-v\tau_0})/v$ and $(v,w)$ are the 
Feynman variational parameters determined by minimizing 
the polaron free energy~\cite{Feynman1955}.
Evaluating this integral numerically at 
$\alpha_h = 2.6$ gives 
$m^*_{\mathrm{pol},h}/m^*_h = 1.6$~\cite{Schultz1959}, 
yielding an effective hole polaron mass 
$m^*_{\mathrm{pol},h} \approx 2.7\,m_0$.
From the fitted plasma frequency, the mobile carrier 
density follows from the Drude relation
\begin{equation}
    n = \frac{\omega_p^2 \varepsilon_0\,
              m^*_{\mathrm{pol},h}}{e^2}
    \approx 5.7 \times 10^{18}\,\mathrm{cm}^{-3},
    \label{eq:carrier_density}
\end{equation}
using $m^*_{\mathrm{pol},h} \approx 2.7\,m_0$ and 
$\omega_p/2\pi = 13$~THz.
Comparing with the maximum photogenerated density
\begin{equation}
    N_\mathrm{max} = \frac{F}{E_\mathrm{ph}\,d}
    \approx 4.0 \times 10^{19}\,\mathrm{cm}^{-3},
    \label{eq:Nmax}
\end{equation}
where $F = 150\,\mu$J\,cm$^{-2}$, 
$E_\mathrm{ph} = 4.64$~eV.

It is important to note that the physical interpretation of Eq.~\eqref{eq:tau_full} depends critically
on the carrier energy relative to the LO-phonon emission threshold
$\hbar\omega_{\mathrm{LO}}$.
Immediately following photoexcitation at $267$~nm
($E_\mathrm{pump} \approx 4.64$~eV), carriers have an excess kinetic
energy $\Delta E \approx 0.40$~eV well above threshold, and the fast phonon-emission channel is active.
We assume, however, that the THz probe at a pump--probe delay of $2.5$~ps samples the carrier population after it has thermalised toward the band  edge through rapid cascaded emission.
At room temperature, the thermal energy
$E_\mathrm{th} = \frac{3}{2}k_BT \approx 39$~meV falls well below
$\hbar\omega_{\mathrm{LO}} \approx 74$~meV, so LO-phonon emission is
kinematically forbidden for the thermalised carrier population \cite{Ridley2013}.
In this near-band-edge regime, the scattering rate is governed solely by LO-phonon absorption, controlled by the Bose-Einstein occupation number
\begin{equation}
    N_{\mathrm{LO}}
    = \frac{1}{\exp\!\left(\hbar\omega_{\mathrm{LO}}/k_BT\right) - 1}.
    \label{eq:NLO}
\end{equation}
For \ch{CeO2} at room temperature this gives $N_{\mathrm{LO}} \approx 0.06$, reflecting the near-absence of thermal phonon emission at this energy scale.
Since $E_k = \tfrac{3}{2}k_BT \approx 39$~meV $
< \hbar\omega_\mathrm{LO}
= 74$~meV, phonon emission is kinematically forbidden and
Eq.~\eqref{eq:tau_full} reduces to the absorption-only limit, with all
terms proportional to $(N_\mathrm{LO}+1)$ set to zero.
Evaluating the remaining absorption terms at the mean thermal energy gives $\tau_m^{\mathrm{LO}} \approx 90 - 96$~fs, which we adopt as the
theoretical Fröhlich-limited momentum relaxation time for thermalised holes in \ch{CeO2} at room temperature~\cite{Ridley2013}.
This gives the corresponding bare mobility of
\begin{equation}
    \mu_\mathrm{bare} = 
    \frac{e\tau_m^\mathrm{LO}}{m^*_{\mathrm{pol},h}}
    \approx 57\,\mathrm{cm^2\,V^{-1}\,s^{-1}}.
    \label{eq:sm_mu_bare}
\end{equation}

\subsubsection{Kadanoff Boltzmann equation mobility}
\label{sm:kadanoff}

The bare Fr\"{o}hlich estimate treats the carrier as 
an independent particle scattered by phonons.
A more complete treatment recognises that the carrier 
is dressed by a cloud of virtual LO phonons, forming 
a polaron whose transport properties differ from those 
of the bare carrier.
The Kadanoff Boltzmann equation 
approach~\cite{Kadanoff1963} solves for the response 
of the Feynman variational 
polaron~\cite{Feynman1955} to an applied field, 
assuming independent scattering events.
In the low-temperature asymptotic limit 
(where $\beta = \hbar\omega_\mathrm{LO}/k_BT \gg 1$ ), the 
Kadanoff mobility is
\begin{equation}
    \mu_K = \left(\frac{w}{v}\right)^3
    \frac{e}{2m^*_h\,\omega_\mathrm{LO}\,\alpha_{h}}\,
    e^{\beta}\,
    \exp\!\left(\frac{v^2-w^2}{w^2 v}\right),
    \label{eq:sm_kadanoff}
\end{equation}
where $(v,w)$ are the Feynman variational parameters,  the factor $(w/v)^3$ accounts for the polaron mass 
renormalization and $\beta = \hbar\omega_\mathrm{LO}/k_BT$ is a reduced thermodynamic temperature
in units of the phonon energy \cite{Frost2017}. 
Evaluating at $\alpha_h = 2.6$ ($v = 3.3$, 
$w = 2.6$) gives 
$\mu_K \approx 15$~cm$^2$\,V$^{-1}$\,s$^{-1}$.

\subsubsection{FHIP path-integral mobility}
\label{sm:fhip}

The FHIP path-integral 
mobility~\cite{Feynman1962} considers the response 
of the polaron centre-of-mass coordinate to a 
dynamically maintained steady state of thermally 
excited phonons.
In the same low-temperature limit, the FHIP mobility 
is related to the Kadanoff result by
\begin{equation}
    \mu_\mathrm{FHIP} = \mu_K \times 
    \frac{3}{2\beta},
    \label{eq:sm_fhip}
\end{equation}
which at $\beta \approx 2.9$ gives 
$\mu_\mathrm{FHIP} \approx 
8$~cm$^2$\,V$^{-1}$\,s$^{-1}$, a full order of 
magnitude below experiment.

We note that both the Kadanoff and FHIP results are 
derived in the low-temperature asymptotic limit 
($\beta \gg 1$); at $\beta \approx 2.9$ their 
quantitative accuracy is limited, and the Hellwarth 
numerical contour 
integration~\cite{Hellwarth1999}, which has been 
applied to polarons in halide 
perovskites~\cite{Frost2017}, would provide a more 
reliable estimate at this temperature.

\subsubsection{Comparison with experiment and evidence from Raman spectroscopy}
\label{sm:mobility_comparison}

Table~\ref{tab:sm_mobility} summarises the three 
theoretical estimates alongside the measured value.
The systematic and substantial underestimation by all 
three models, which span progressively more 
sophisticated treatments of the polaron--phonon 
interaction, is striking.

\begin{table}[h]
    \centering
    \caption{Comparison of theoretical polaron 
    mobility estimates with the experimentally 
    measured value (100~cm$^2$/Vs) for holes in photoexcited 
    \ch{CeO2} at 295~K.}
    \label{tab:sm_mobility}
    \begin{tabular}{lcc}
        \toprule
        Model & $\mu$ 
        (cm$^2$\,V$^{-1}$\,s$^{-1}$) 
              & $\mu/\mu_\mathrm{exp}$ \\
        \midrule
        FHIP~\cite{Feynman1962}      
              & 10  & 0.10 \\
        Kadanoff~\cite{Kadanoff1963} 
              & 20  & 0.18 \\
        Bare Fr\"{o}hlich            
              & 57  & 0.55 \\
%        \midrule
%        \textbf{Experiment (Drude fit)} 
%              & \textbf{103} & \textbf{1.00} \\
        \bottomrule
    \end{tabular}
\end{table}

\begin{figure}[!t]
    \centering
    \includegraphics[width=\columnwidth]{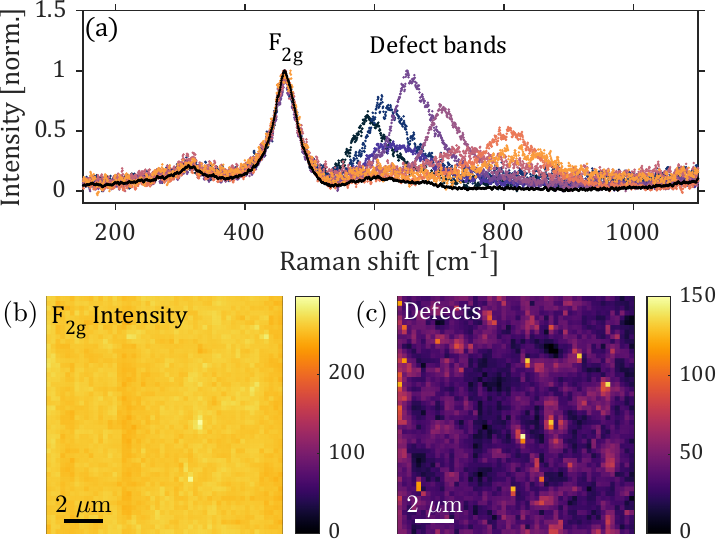}
    \caption{\textbf{(a)} Raman spectra acquired using a 532~nm laser excitation on a 50~nm CeO$_2$ thin film supported by a diamond substrate, following UV photoexcitation. \textbf{(black)} Representative mean Raman spectrum of the CeO$_2$ film with the expected $F_{2g}$ mode at 462~cm$^{-1}$ and a surface phonon feature at 314~cm$^{-1}$. A broad phonon contribution centered near 607~cm$^{-1}$ is assigned to oxygen-vacancies and associated with Ce$^{3+}$-like defects. \textbf{(color)} Local anomalous Raman spectra exhibiting enhanced defect-related features. Defect signatures near 600~cm$^{-1}$ are consistent with conventional oxygen-vacancy-related modes, while higher-frequency peaks are likely linked to adsorbed oxygen/peroxide-like surface species. \textbf{(b)}, \textbf{(c)} 12$\times$12$~\mu$m$^2$ Raman maps of the $F_{2g}$ peak intensity and integrated defect intensity, respectively, with the defect response integrated from 530~cm$^{-1}$ to 1000~cm$^{-1}$. A broad integration window was chosen owing to the distinct defect feature frequencies present across the sample. The integrated defect intensity exhibits strong nonuniformity. This spatial clustering of the defects, together with the broad frequency range spanned by these modes, provides evidence of localized structural modification of the CeO$_2$ following UV photoexcitation.}
    \label{fig:RamanSpectrum}
\end{figure}

The discrepancy is deepened by the Raman data.
By Matthiessen's rule \cite{Ashcroft1976}, any additional elastic scattering from structural defects can only add to the total scattering rate,
\begin{equation}
    \frac{1}{\tau_\mathrm{tot}}
    = \frac{1}{\tau_\mathrm{LO}}
    + \frac{1}{\tau_\mathrm{def}}
    + \cdots,
    \label{eq:matthiessen}
\end{equation}
so the presence of disorder should make $\tau_\mathrm{tot}$ 
\emph{shorter} than $\tau_m^\mathrm{LO}$, not longer. Raman spectroscopy confirms that such disorder is indeed present 
in the film. Raman spectroscopy of the film (Fig.~\ref{fig:RamanSpectrum}a) reveals an expected fluorite structure $F_{2g}$ mode at 462~cm$^{-1}$ and an additional peak near  314~cm$^{-1}$ attributed to surface phonon modes in CeO$_2$ thin films~\cite{WANG2001246}. Phonon bands between 530~cm$^{-1}$ and 600~cm$^{-1}$ are associated with oxygen vacancies and Ce$^{3+}$ defect sites~\cite{McBride1994,Schmitt2020,Xu2019}, and the presence of higher frequency bands likely reflects oxygen-related surface species, such as adsorbed oxygen or peroxide-like defects, localized at strongly modified regions of the CeO$_2$ surface ~\cite{Schilling2017,Wu2010}. A spatially resolved Raman map of the $F_{2g}$ intensity (Fig.~\ref{fig:RamanSpectrum}b) reveals a homogeneous film, while the map of the integrated defect intensity (Fig.~\ref{fig:RamanSpectrum}c) shows strong spatial variations. These enhanced defect features are present in the region exposed to prolonged UV irradiation, indicating photoinduced degradation of the film during the course of the THz measurements, confirmed by the absence of these defect bands in the Raman spectra of the region that was not exposed to UV irradiation. 
The film therefore contains a substantial population of point 
defects that act as additional elastic scattering centres, yet 
the observed $\tau_\mathrm{Drude}$ exceeds the bare phonon-limited value by a factor of two. This discrepancy cannot be reconciled within the bare Fr\"{o}hlich 
picture and points to a breakdown of the independent-particle 
description of transport.

We interpret this observation in the framework of large-polaron 
physics at intermediate coupling strength ($\alpha_h = 2.5$--$2.8$). In this regime, the charge carrier is not a bare 
quasiparticle but a polaron, a carrier dressed by a cloud of 
virtual LO phonons, and the Drude model, when applied to the 
polaron optical conductivity, does not extract the bare phonon scattering time but rather an effective relaxation time of the polaronic quasiparticle~\cite{Mishchenko2019}. In this  intermediate-coupling regime, the dressing of charge carriers  by LO phonons reduces momentum scattering while simultaneously  increasing the carrier effective mass~\cite{Lan2019}, and polaron 
formation has been shown to suppress scattering with phonons, other carriers, and charged defects alike~\cite{Bretschneider2018}. 
The observation that $\tau_{\mathrm{Drude}} \gg \tau_m^{\mathrm{LO}}$ 
even in the presence of the oxygen-vacancy and Ce$^{3+}$ defects 
evidenced by the Raman spectrum is therefore consistent with the 
polaronic suppression of momentum scattering, and inconsistent 
with the bare Fr\"{o}hlich picture in which defect scattering 
would only further reduce $\tau$ via Matthiessen's 
rule~(Eq.~\ref{eq:matthiessen}). A quantitative account of the 
effective relaxation time within the Feynman polaron 
model~\cite{Feynman1955, Mishchenko2019} is left for future work.

\subsection{Density functional theory calculations}
Electronic properties of \ch{CeO2} were calculated using 
density functional theory (DFT) as implemented in the Vienna 
\textit{ab initio} Simulation Package (VASP) version 
6.5.1~\cite{Kresse1993, Kresse1994, Kresse1996Efficiency, 
Kresse1996Efficient}. The projector-augmented wave (PAW) 
potentials~\cite{Blochl1994, 
Kresse1999} were used for both Ce and O, treating the 
Ce~$5s$, $5p$, $6s$, $5d$ and $4f$ states as valence 
electrons (12 valence electrons for Ce) and the O~$2s$ and 
$2p$ states as valence electrons (6 valence electrons for O).
A plane-wave energy cutoff of 550~eV was used 
throughout.
Two different exchange-correlation functionals were 
employed, chosen according to the balance between 
accuracy and computational cost required by each 
calculation.
The electronic band structure was computed on the primitive fluorite unit cell (3 atoms: 1~Ce + 2~O) using the hybrid functional of Perdew, Ernzerhof, and Burke (PBE0)~\cite{Perdew1996,Adamo1999}, which incorporates 25\% Hartree--Fock exchange.
\begin{comment}
The electronic band structure was computed on the 
primitive fluorite unit cell (3 atoms: 1~Ce + 2~O) 
using the screened hybrid functional of Heyd, Scuseria, 
and Ernzerhof (HSE06)~\cite{Heyd2003}, which 
incorporates 25\% screened Hartree--Fock exchange with a 
screening parameter $\mu = 0.2$~\AA$^{-1}$. 
\end{comment}

Structural parameters were taken from the experimental 
lattice constant $a = 5.411$~\AA~\cite{Kummerle1999} 
and subject to structural relaxation.
A $\Gamma$-centred $6\times6\times6$ $k$-point mesh was 
used for the self-consistent calculation, followed by a 
non-self-consistent band structure evaluation along the 
high-symmetry path 
$\Gamma$-$\text{X}$-$\text{W}/\text{K}$-$\text{T}$-$\text{L}$-$\text{U}$-$\text{W}$-$\text{L}$-$\text{K}$, 
with an electronic convergence criterion of $10^{-6}$~eV.
Born effective charges and the high-frequency dielectric 
tensor were computed using density functional 
perturbation theory 
(DFPT)~\cite{Baroni1987,Gonze1997} on a 
$2\times2\times2$ supercell of the conventional fluorite 
unit cell (96 atoms: 32~Ce + 64~O), using the GGA-PBE 
exchange-correlation functional~\cite{Perdew1996} with 
an on-site Hubbard correction on the Ce~$4f$ orbitals 
in the simplified rotationally invariant formulation of 
Dudarev~\textit{et al.}~\cite{Dudarev1998}.
Two values of $U_\mathrm{eff}$ were used at different 
stages of the workflow.
Structural relaxation of both the pristine conventional 
cell and the $2\times2\times2$ supercell was performed 
with $U_\mathrm{eff} = 4.5$~eV, the standard value for 
reproducing bulk ground-state properties of 
\ch{CeO2}~\cite{Fabris2005,Da2007}.
All subsequent electronic structure and DFPT 
calculations, on both the pristine and polaron 
supercells, were performed with 
$U_\mathrm{eff} = 6.0$~eV.
The higher value was adopted because it is required to 
stabilise a localised Ce$^{3+}$ polaron in the 
$2\times2\times2$ 
supercell~\cite{Castleton2019}; using the same 
$U_\mathrm{eff}$ for both the pristine and polaron 
DFPT ensures that the difference in Born effective 
charges $\Delta Z^*$ reflects the effect of polaron 
formation alone, rather than a change in functional.

For the polaron supercell, an extra electron was added 
($N_\mathrm{elect} = N_\mathrm{neutral} + 1 = 769$) 
and localised on a single Ce site by displacing the 
eight nearest-neighbour oxygen atoms outward along 
their Ce--O bond directions by an amount corresponding 
to 50\% of the Shannon ionic radius expansion for the 
Ce$^{4+}$ $\to$ Ce$^{3+}$ 
conversion~\cite{Shannon1969}, combined with a magnetic 
moment seed on the target site.
The geometry was relaxed with 
$U_\mathrm{eff} = 6.0$~eV and fixed cell parameters 
until forces converged below 0.01~eV/\AA, yielding a 
total magnetic moment of 
$\sim\!1.0\,\mu_\mathrm{B}$ localized on the 
Ce$^{3+}$ site.
Symmetry was fully disabled for all 
polaron calculations, since the Ce$^{3+}$ site breaks 
the cubic symmetry of the pristine cell.

\subsubsection{Born Effective Charges and Oscillator Strength}
\label{appendix:oscillator_strength}
The oscillator strength of an infrared-active phonon 
mode is determined by the Born effective charges of the 
ions participating in the mode.
Starting from the ionic contribution to the macroscopic 
polarization and the equation of motion of the $F_{1u}$ 
phonon mode driven by a time-varying electric field at 
frequency $\omega$, the dielectric function can be 
written as~\cite{Gonze1997,Baroni2001}
\begin{equation}
    \varepsilon(\omega) = \varepsilon_\infty 
    + \frac{e^2}{\varepsilon_0\,\Omega}
    \sum_j \frac{\left(Z^*_{\mathrm{eff},j}\right)^2}
    {\omega^2_{\mathrm{TO},j} - \omega^2},
    \label{eq:eps_bec}
\end{equation}
where $\Omega$ is the unit cell volume and 
$Z^*_{\mathrm{eff},j}$ is the mode-effective dynamical 
charge, obtained by projecting the site-resolved Born 
effective charge tensors onto the phonon eigenvector:
\begin{equation}
    Z^*_{\mathrm{eff},j} 
    = \sum_\kappa 
      \frac{\mathbf{Z}^*_\kappa \cdot 
            \boldsymbol{\xi}^{(j)}_\kappa}
           {\sqrt{m_\kappa}},
    \label{eq:mode_bec_appendix}
\end{equation}
where $\boldsymbol{\xi}^{(j)}_\kappa$ is the 
mass-weighted eigenvector of mode $j$, normalised such 
that 
$\sum_\kappa |\boldsymbol{\xi}_\kappa|^2 = 1$ and 
$\sum_\kappa m_\kappa \boldsymbol{\xi}_\kappa 
= \mathbf{0}$.

Comparing Eq.~\eqref{eq:eps_bec} with the dimensionless 
Lorentz oscillator form 
$\varepsilon(\omega) = \varepsilon_\infty + 
\sum_j S_j\,\omega^2_{\mathrm{TO},j}/(\omega^2_{\mathrm{TO},j} - \omega^2)$, 
the oscillator strength of mode $j$ is identified as
\begin{equation}
    S_j = \frac{e^2 
    \left(Z^*_{\mathrm{eff},j}\right)^2}
    {\varepsilon_0\,\Omega\,
    \omega^2_{\mathrm{TO},j}}.
    \label{eq:oscillator_bec_appendix}
\end{equation}
A more detailed derivation of this relationship and the 
mode-resolved ionic contributions to the static dielectric 
tensor can be found in the foundational density functional 
perturbation theory literature~\cite{Gonze1997,Baroni2001}.
Taking the logarithmic derivative yields the oscillator 
strength budget equation used in the main text:
\begin{equation}
    \frac{\Delta S}{S} \approx 
    2\frac{\Delta Z^*_\mathrm{eff}}{Z^*_\mathrm{eff}} 
    - \frac{\Delta\Omega}{\Omega} 
    - 2\frac{\Delta\omega_\mathrm{TO}}
            {\omega_\mathrm{TO}},
    \label{eq:deltaS_appendix}
\end{equation}
where the three terms represent the contributions from 
changes in the mode-effective charge, the unit cell 
volume, and the TO frequency, respectively.
At the photoexcited carrier densities studied in this 
work, $\Delta\Omega/\Omega \approx 0$.

Table~\ref{tab:bec_comparison} summarises the 
site-resolved Born effective charges from the DFPT 
calculations.
The acoustic sum rule 
$\sum_\kappa Z^*_\kappa = 0$ is satisfied to within 
$10^{-4}\,|e|$ for the pristine cell and 
$10^{-5}\,|e|$ for the polaron cell.

\begin{table}[t]
    \centering
    \caption{Site-resolved isotropic Born effective 
    charges $Z^* = \mathrm{Tr}(\mathbf{Z}^*)/3$ 
    (in units of $|e|$) from DFPT calculations on the 
    $2\times2\times2$ supercell of \ch{CeO2}. The 
    polaron site is the Ce atom on which the extra 
    electron is localised.}
    \label{tab:bec_comparison}
    \begin{tabular}{lccc}
        \toprule
        Site & Pristine & Polaron & $\Delta Z^*$ \\
        \midrule
        Ce (polaron site)            & $+5.63$ & $+4.35$ & $-1.28$ \\
        Ce (far, avg.\ of 31)        & $+5.63$ & $+5.57$ & $-0.06$ \\
        O (nearest, avg.\ of 8)      & $-2.82$ & $-2.80$ & $+0.01$ \\
        O (far, avg.\ of 56)         & $-2.82$ & $-2.76$ & $+0.05$ \\
        \midrule
        $|\sum_\kappa Z^*_\kappa|$ 
                & $<10^{-13}$ & $<10^{-5}$ & \\
        \bottomrule
    \end{tabular}
\end{table}
Projecting the BEC tensors onto the $F_{1u}$ phonon 
eigenvector gives mode-effective charges of 
$\tilde{Z}^*_{F_{1u}} = 6.15$~$e\cdot
\mathrm{amu}^{-1/2}$ (pristine) and 
$6.02$~$e\cdot\mathrm{amu}^{-1/2}$ (polaron), a 
fractional change of 
$\Delta\tilde{Z}^*/\tilde{Z}^* \approx -2.0\%$. 
Via Eq.~\eqref{eq:deltaS_appendix}, this gives 
$\Delta S/S \approx -3.99\%$ at the supercell polaron 
concentration of $1/32$. We note that the sign and order of magnitude are consistent with the 
experimentally observed $\Delta S/S \approx -2$ to 
$-3\%$, although a direct numerical comparison requires 
care as the DFPT calculation represents a static 
equilibrium configuration at a different polaron 
fraction than the experiment.

\subsection{Exciton Binding Energy and Saha Analysis}
\label{sm:saha}

The equilibrium free-carrier fraction 
$\xi = n_\mathrm{free}/N_\mathrm{max}$ in the 
presence of exciton formation is governed by the 
Saha--Langmuir equation~\cite{Saha1921}
\begin{equation}
    \frac{\xi^2}{1 - \xi}
    = \frac{1}{N_\mathrm{max}}
      \left(\frac{\mu_X k_B T}{2\pi\hbar^2}
      \right)^{\!3/2}
      e^{-E_b/k_BT},
    \label{eq:sm_saha}
\end{equation}
where 
$\mu_X = m^*_e m^*_h/(m^*_e + m^*_h) 
\approx 1.29\,m_0$ is the exciton reduced mass and 
$E_b$ is the hydrogenic exciton binding energy
\begin{equation}
    E_b = \frac{\mu_X}{m_e}
          \frac{\mathrm{Ry}}{\varepsilon^2},
    \label{eq:sm_Eb}
\end{equation}
with $\mathrm{Ry} = 13.6$~eV and $\varepsilon$ the 
dielectric constant screening the electron--hole 
Coulomb interaction.

The appropriate choice of $\varepsilon$ depends on the 
ratio of $E_b$ to the LO phonon energy 
$\hbar\omega_\mathrm{LO}$~\cite{Haken1959}.
When $E_b \gg \hbar\omega_\mathrm{LO}$, the electron 
orbits faster than the lattice can respond and only 
the electronic polarization screens the interaction 
($\varepsilon_\infty$); when 
$E_b \ll \hbar\omega_\mathrm{LO}$, the lattice fully 
relaxes and the static constant $\varepsilon_s$ 
applies. 
The $\varepsilon_\infty$-screened binding energy 
($E_b \approx 623$~meV) gives an exciton Bohr radius 
of only $\sim\!2.2$~\AA, smaller than the lattice 
constant $a = 5.41$~\AA, invalidating the 
Wannier--Mott continuum model on which 
Eq.~\eqref{eq:sm_Eb} is based.
The $\varepsilon_s$-screened value 
($E_b \approx 30$~meV) predicts $\xi \approx 83\%$, 
meaning that most carriers would remain free, which is far 
more than the $14\%$ observed experimentally.

Since $E_b$ falls in the intermediate regime 
$E_b \sim \hbar\omega_\mathrm{LO}$ for any plausible 
choice of screening, we adopt the geometric-mean 
interpolation 
$\varepsilon_\mathrm{eff} = 
\sqrt{\varepsilon_\infty\,\varepsilon_s} 
\approx 11.5$, which approximates the crossover 
between electronic and ionic screening in the Haken 
potential~\cite{Haken1959}.
This gives $E_b \approx 132$~meV and an exciton Bohr 
radius $a_X \approx 4.8$~\AA, comparable to the 
lattice constant of \ch{CeO2}.
Evaluating Eq.~\eqref{eq:sm_saha} at $T = 295$~K 
yields $\xi \approx 0.23$, indicating that 
approximately $77\%$ of photogenerated pairs form 
bound excitons, reducing the free-carrier pool to 
$n_\mathrm{free} \approx 9.2 \times 
10^{18}$~cm$^{-3}$, in reasonable agreement with the 
$n \approx 5.7 \times 10^{18}$~cm$^{-3}$ extracted 
from the Drude fit.

We note that this agreement should be viewed as 
qualitative rather than quantitative, given the 
exponential sensitivity of 
Eq.~\eqref{eq:sm_saha} to $E_b$ and the inherent 
uncertainty in the effective screening.
Furthermore, carrier trapping at grain boundaries and point defects in the polycrystalline film may also contribute to the reduction of the mobile carrier fraction, independently of exciton formation.

\nocite{*}

\bibliography{apssamp2}% Produces the bibliography via BibTeX.
\bibliographystyle{apsrev4-2.bst}
\end{document}